\documentclass[twocolumn,superscriptaddress,reprint]{revtex4-1}

\usepackage[pdftex]{graphicx}   
\usepackage{nicefrac}
\usepackage{xcolor}
\usepackage{subfigure}
\usepackage{amsmath}

\usepackage[section]{placeins} 
\usepackage{wrapfig}

\newcommand{\up}{{\uparrow}}
\newcommand{\dw}{{\downarrow}}
\newcommand{\pd}{{\phantom{\dagger}}}

\begin{document}

\title{Correlation-Driven Charge Order in a Frustrated Two-Dimensional Atom Lattice}

\author{F. Adler}
\affiliation{Physikalisches Institut and W\"urzburg-Dresden Cluster of Excellence ct.qmat, Universit\"at W\"urzburg, D-97074 W\"urzburg, Germany}
\author{S. Rachel}
\affiliation{School of Physics, University of Melbourne, Parkville, VIC 3010, Australia}
\affiliation{Institut f\"ur Theoretische Physik, Technische Universit\"at Dresden, D-01069 Dresden, Germany}
\author{M. Laubach}
\affiliation{Institut f\"ur Theoretische Physik, Technische Universit\"at Dresden, D-01069 Dresden, Germany}
\author{J. Maklar}
\affiliation{Physikalisches Institut and W\"urzburg-Dresden Cluster of Excellence ct.qmat, Universit\"at W\"urzburg, D-97074 W\"urzburg, Germany}
\author{A. Fleszar}
\affiliation{Institut f\"ur Theoretische Physik und Astrophysik, Universit\"at W\"urzburg, D-97074 W\"urzburg, Germany}
\author{J. Sch\"afer*}
\affiliation{Physikalisches Institut and W\"urzburg-Dresden Cluster of Excellence ct.qmat, Universit\"at W\"urzburg, D-97074 W\"urzburg, Germany}
\author{R. Claessen}
\affiliation{Physikalisches Institut and W\"urzburg-Dresden Cluster of Excellence ct.qmat, Universit\"at W\"urzburg, D-97074 W\"urzburg, Germany}


\begin{abstract}
We thoroughly examine the ground state of the triangular lattice of Pb on Si(111) using scanning tunneling microscopy and spectroscopy.
We detect electronic charge order, and disentangle this contribution from the atomic configuration which we find to be 1-down---2-up, contrary to previous predictions from density functional theory.
Applying an extended variational cluster approach we map out the phase diagram as a function of local and nonlocal Coulomb interactions. Comparing the experimental data with the theoretical modeling leads us to conclude that electron correlations are the driving force of the charge-ordered state in Pb/Si(111).
These results resolve the discussion about the origin of the well-known $3\times 3$ reconstruction.
By exploiting the tunability of correlation strength, hopping parameters, and bandfilling, this material class represents a promising platform to search for exotic states of matter, in particular, for chiral topological superconductivity.
\end{abstract}

\maketitle


In a frustrated lattice of uncompensated spins the exchange interactions cannot be saturated completely on every site.
This leads to competing ground states where either a specific magnetic order or a spin liquid phase can emerge\,\cite{huse-88prl2531, nakatsuji2005, balents2010, zhu-15prb041105, savary-16rpp016502}.
When nonlocal Coulomb interactions are involved or the system is doped away from half filling, the formation of charge-order (CO) is another possibility. These scenarios are often accompanied by superconductivity arising in the vicinity of such ordered phases.
Yet, candidate materials are limited to very few bulk solids, such as cobaltates\,\cite{takada2003, baskaran03prl097003} and organic compounds \,\cite{
kanoda1997, shimizu2003}. However, due to the complexity of these materials, the occurrence of particular phases is not fully understood.
In contrast, atomic two-dimensional (2D) lattices with a triangular net, experimentally generated by epitaxial submonolayer deposition on an insulating substrate, are intriguingly simple in structure. Thus they provide versatile model systems for the study of strong electron correlations.
The generically rich phase diagram of correlated triangular systems has been pointed out in theoretical studies of lattice models \cite{motrunich2004, watanabe2005} and surface systems \,\cite{PhysRevLett.98.086401, BiermannGWDMFT}, including CO and the possibility of topological superconductivity\,\cite{mackenzie-03rmp657,vojta1999, 
Weber2006,nandkishore-12np158,kiesel2013,sato-17rpp076501,Cao2017}. In this respect, the atomic architecture allows us to tune the interactions by variation of the adatom species as well as the substrate which provides screening and mediates the electron hopping \,\cite{GlassPRL2015}. In addition, dopants such as alkali atoms have been demonstrated to change the band filling \,\cite{HiraharaPRB2009}.

\begin{figure}[htb]
\centering
\includegraphics[width=\linewidth]{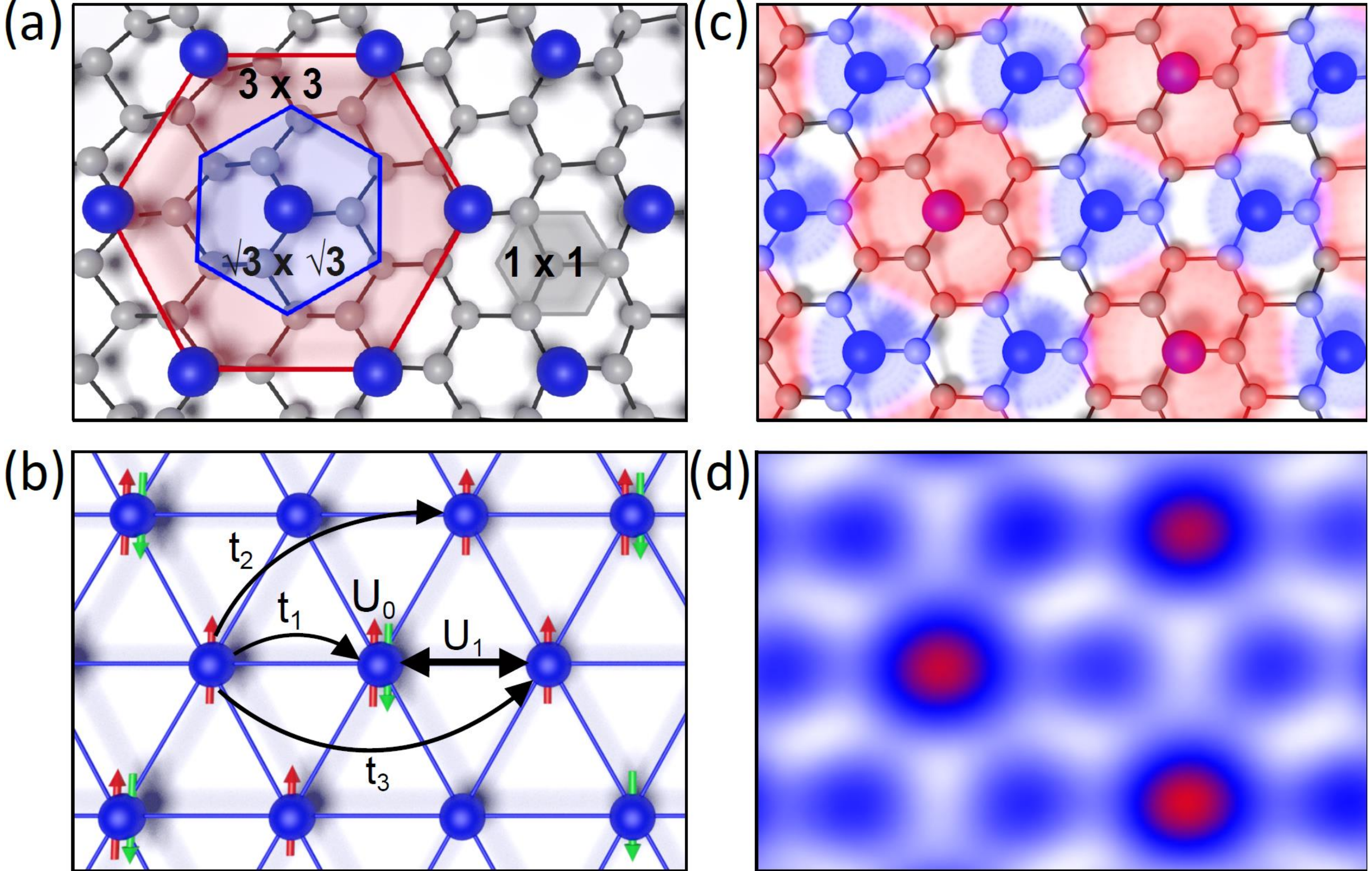}
\caption{(a) Structure model of a $\sqrt{3}\times\sqrt{3}$ adatom lattice on a Si(111) substrate. The Wigner-Seitz unit cells of the substrate surface (gray), the adatom lattice (blue) and the charge-ordered state (red) are depicted; (b) Illustration of the involved parameters governing the different ground states: hopping integrals $t_1$, $t_2$, $t_3$, local Coulomb interaction $U_0$ and nearest-neighbor Coulomb interaction $U_1$; (c) Model representation of Pb/Si(111) in the $3 \times 3$ charge-ordered state. Red and blue charge clouds reflect the excess and reduced charge density at the respective adatom sites. (d) Experimental STM data of Pb/Si(111) showing the charge-ordered state.}
\label{fig:teaser}
\end{figure}

The case in point are group-IV adsorbates (Sn, Pb) on semiconductor surfaces such as Si, Ge, or SiC\,\cite{PhysRevLett.98.086401, PbSiSTMintrinsic, GlassPRL2015}. The key concept here is that unsaturated adsorbate orbitals exist with half filling which are subject to significant local and nonlocal Coulomb interactions. These surface systems thus represent a rich playground for the investigation of correlation physics in a frustrated lattice, including the formation of unusual symmetry-broken ground states.
The experimental system is a triangular array of atoms with a dilute coverage of a 1/3 monolayer, forming a $\sqrt{3}\times\sqrt{3}$ surface reconstruction [Fig.\,\ref{fig:teaser}(a)] where the adsorbed Sn or Pb atoms are known to reside in the  $T_4$ position\,\cite{PbSiT4LDA, SnSiT4XSW}. This implies that three out of their four valence orbitals are engaged in covalent back bonds to the substrate. Of relevance for the physics is the fourth orbital (out-of-plane $p_z$ orbital): it remains ``dangling'' and contains only one electron. Such a half-filled surface band is prone to significant electron correlations, and here the on-site Coulomb repulsion becomes a relevant term due to the weak hopping matrix elements, as illustrated in Fig.\,\ref{fig:teaser}(b). At low temperatures, some systems undergo a phase transition to a $3 \times 3$ superstructure. For Sn and Pb on Ge(111) it was initially interpreted as a Peierls distortion\,\cite{PbGeNatureCDW, SnGeCDW}. However, subsequently this was heavily debated due to insufficient nesting conditions in the electron band structure\,\cite{MascaraquePRB98}. With the obvious presence of non-negligible electron correlations, a single-particle description must appear insufficient\,\cite{CDWMottHartreeFock}, calling for approaches that account for the relevant many-body interactions.

\begin{figure}[htbp]
\centering
\includegraphics[width=\linewidth]{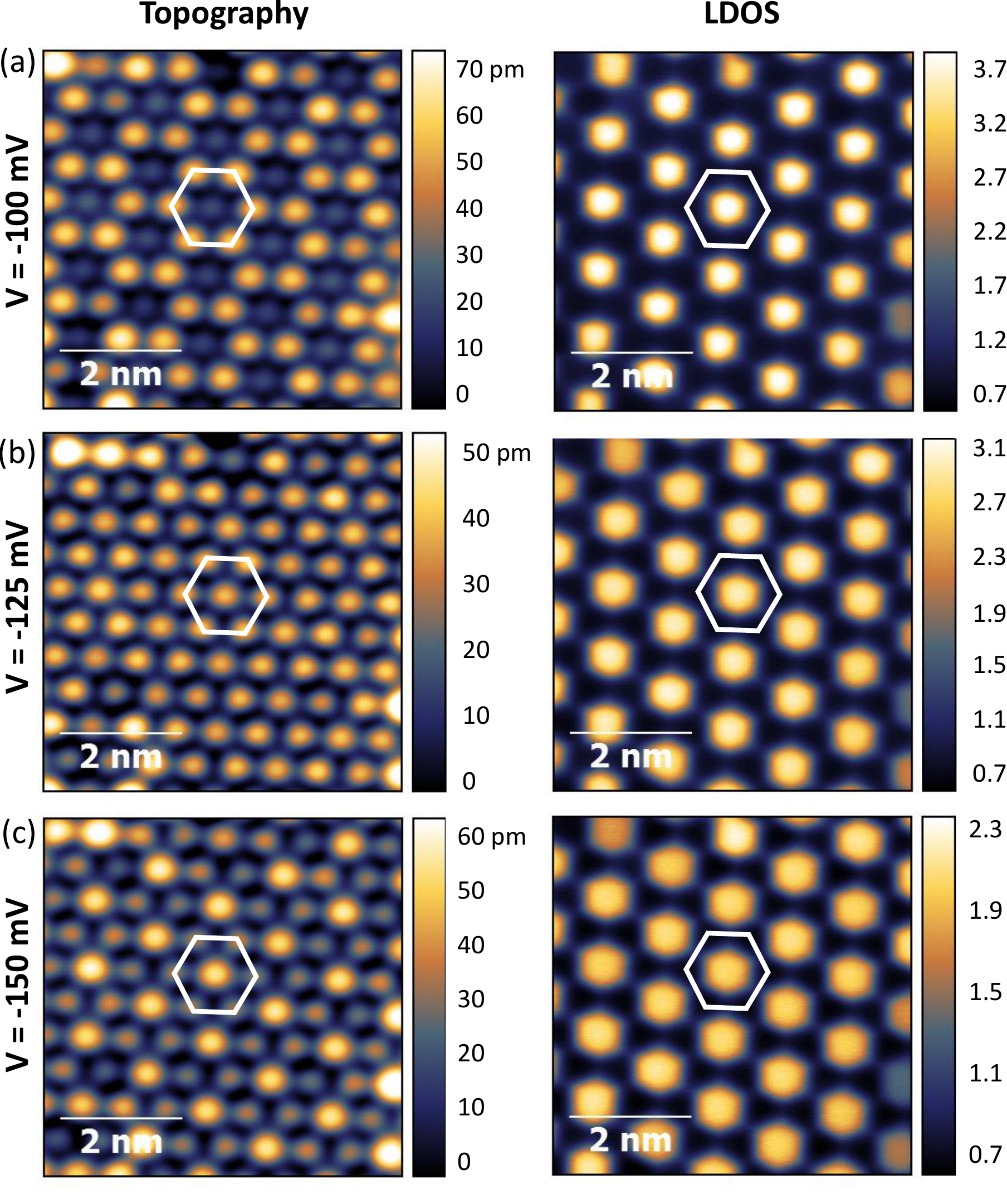}
\caption{Topography (left) and the corresponding LDOS maps (right) of Pb/Si(111) for bias voltages (a) $V=-100$, (b) $V=-125$, and (c) $V=-150$\,mV. A $3 \times 3$ Wigner-Seitz unit cell is marked in each panel. While the LDOS maps show the CO state with a qualitatively unchanged appearance, the contrast in the topographic maps switches due to the interplay of LDOS effects from the CO state with the concomitant topographic buckling. The modulation voltage used to record the LDOS maps was 10\,mV at a frequency of 653\,Hz.}
\label{fig:exp}
\end{figure}

Pb on Si(111) might be the experimentally least studied system of this kind, since it coexists with other surface phases, see Supplemental Material \cite{sup}, thereby prohibiting the use of spatially averaging techniques such as photoemission. The symmetry-broken ground state with $3 \times 3$ superstructure, Fig.\ \ref{fig:teaser}(c), exists below 86\,K\,\cite{PbSiSTM3x3, PbSiSTMintrinsic}, and lacks a clear explanation so far.
Sources for this reconstruction could be a simple structural transition or an effect of the electronic system, i.e.\ {\it charge ordering}. The latter can be due to a Peierls-type charge density wave (CDW) instability\,\cite{PbGeNatureCDW, SnGeCDW} caused by Fermi surface nesting, or can be induced by longer-ranged electron-electron interactions\,\cite{BiermannGWDMFT}. In principle such a correlation-driven CO might compete with other, possibly magnetically ordered phases, as observed in Sn/Ge(111) \cite{Cortes}.
Moreover, a recent study of Pb/Si(111) explored the role of spin-orbit coupling\,\cite{Cren}.

In this Letter, we investigate Pb/Si(111) using a local probe, namely low-temperature scanning tunneling spectroscopy (STS), which is sensitive to the local density of states (LDOS). Thereby we are able to identify CO [Fig.\ \ref{fig:teaser}(d)] as the ground state of the system, which we can clearly distinguish from an accompanying weak structural distortion.
Using quasiparticle interference (QPI) we determine the dispersion of the spectral function in the vicinity of the Fermi level. The QPI data agree well with many-body simulations using the extended variational cluster approach (XVCA) where long-ranged electron-electron interactions turn out to be crucial. Pb/Si(111) is found to be located in the metallic CO regime of the phase diagram. The presence of nearest-neighbor Coulomb repulsion on the triangular lattice leads to frustration of the charge configuration.

\emph{Experimental and theoretical methods.---}
For preparation we used $p$-doped Si(111) substrates with a sheet resistance $\rho <0.02\,\Omega$cm to ensure sufficient conductivity for tunneling at low temperatures. The substrates were prepared by repeated flashing at $T=1650$\,K for 10\,s followed by a slow cooldown to room temperature until a sharp $7 \times 7$ LEED pattern was observed.
The Pb/Si(111) surface was prepared by evaporating $\sim 1$\,ML of Pb on a substrate kept at room temperature followed by an annealing step at $T=720$\,K for 5\,min. This results in a surface which, at room temperature, is mostly covered with $1 \times 1$ Pb/Si(111) and small islands of the desired $\sqrt{3}\times\sqrt{3}$ reconstruction which was in the focus of this study.
Our experiments were performed with an Omicron LT-STM with a base pressure $<5 \times 10^{-11}$\,mbar at $T=4.3$\,K, i.e.,\ deep in the low-temperature $3 \times 3$ phase\,\cite{PbSiSTM3x3, PbSiSTMintrinsic}. The utilized tungsten tips were tested on a Ag(111) single crystal for sharpness and spectroscopic properties. All STM data shown here were recorded at constant tunneling current with set point $I=200$\,pA. LDOS maps were recorded using a modulation spectroscopy technique with a lock-in amplifier.

The XVCA method extends the variational cluster approach (VCA)\,\cite{Potthoff2003b,Potthoff2003a} originally introduced for Hubbard models with local interactions only. This extension takes into account longer-ranged interactions\,\cite{Aichhorn2004a}. Both XVCA and VCA provide an efficient scheme to approximately solve the many-body problem, with the single-particle Green's function $\mathcal{G}(\mathbf{k},\omega)$ as the central outcome.
Using $\mathcal{G}(\mathbf{k},\omega)$, we can directly compute LDOS and the single-particle spectral function $\mathcal{A}(\mathbf{k},\omega)$. A detailed description and discussion of the XVCA method is given in the Supplemental Material \cite{sup}.

\emph{Disentangling charge order and structural distortion.---} Figure \ref{fig:exp} shows topographic images and the corresponding LDOS maps for selected tunneling biases, all recorded at the exact same location on the sample. In the LDOS maps, we observe a regular large-scale redistribution of the charge over the spatial coordinates, where the atom in the center of the hexagonal Wigner-Seitz (WS) unit cell shows a significantly enhanced LDOS, while the six surrounding atoms in the corner fade into the background and can hardly be resolved individually. Such characteristic behavior is found for a wide range of other sample biases, see Fig. S3 in the Supplemental Material \cite{sup}.
As illustrated in Fig.\ \ref{fig:teaser}(c), charge is accumulated on 1/3 of the adatoms, while it is depleted on the neighboring ones.
This experiment thereby  manifests CO in the Pb atom lattice on Si(111), and in conjunction with the theoretical modeling below this is established as driven by correlations.

In looking at the topographic images in Fig.\ \ref{fig:exp}, one notes a qualitative change of the pattern with increasing bias. Here, one needs to take into account that topographic images suffer from an imprinting of LDOS contributions. The recorded apparent ``height'' of an atom at given bias is not only determined by the topographic corrugation, but also by the energy-integrated LDOS (from the Fermi energy to the energy that corresponds to the bias). This effect can be disentangled in a straightforward manner by analyzing the bias-dependent signal.

Starting at small negative bias of $V \geq -100$\,mV [Fig.\ \ref{fig:exp}(a)] the atom in the center of the WS unit cell appears lower than neighboring atoms. Since there is only little contribution from the integrated LDOS at low bias voltages, this qualitatively reflects the true corrugation of the sample surface, i.e., Pb atoms are arranged in a ``1-down---2-up'' configuration. The total height difference we observe between ``up'' and ``down'' atoms is only 67\,pm \cite{sup}. With increasing absolute bias voltage the CO state progressively imprints its charge distribution onto the topographic maps, raising the apparent height of the atom in the center of the WS unit cell. Eventually, for $V \leq -150$\,meV, this atom appears higher than surrounding atoms, see Fig.\ \ref{fig:exp}(c), rendering the topographic maps completely dominated by LDOS effects.
Thus, there is a weak topographic lattice distortion with a specific atomic pattern which is out of phase with the CO pattern. These findings are in contrast to predictions from density functional theory (DFT), which favor a ``1-up---2-down'' configuration \cite{PbSiSTMintrinsic, Cren}. However, this is not surprising since it was already shown for Sn/Ge(111) that DFT results are sensitive to the inclusion of electronic correlations \cite{PhysRevLett.98.086401}, pointing towards the limited capability to correctly describe these correlated systems on a DFT level.

\begin{figure*}[htbp]
  \centering
  \includegraphics[width=\textwidth]{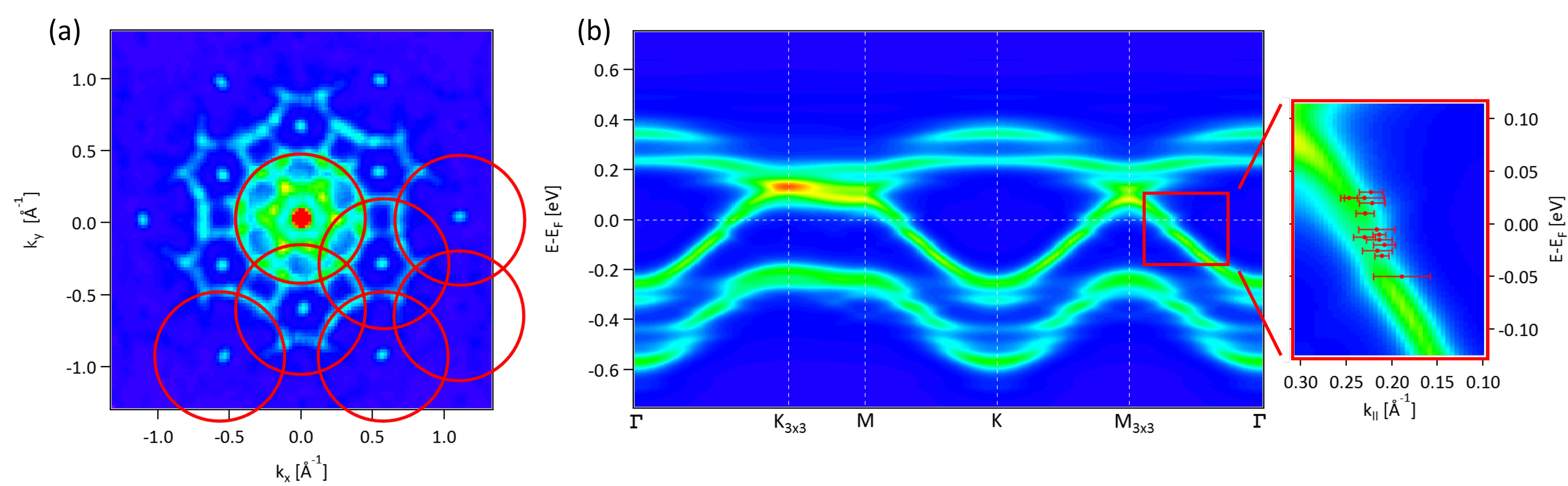}
  \caption{(a) Quasiparticle interference pattern obtained from Fourier-transforming a $28 \times 28\,\textrm{nm}^2$ $dI/dV$ map of Pb/Si(111) measured at $V=+10$\,mV (modulation voltage 2.5\,mV, image symmetrized with respect to the threefold symmetry, three-point Gaussian smoothing). The seemingly complex pattern can be explained by a single scattering channel as indicated by the red circles around selected Bragg spots.
  (b) Theoretically calculated $k$-resolved spectral function of Pb/Si(111) for parameters $(U_0/t_1, U_1/t_1) =(5,3)$. The Fermi vector $k_f \approx 0.22\,$\r{A}$^{-1}$ is in good agreement with the scattering vector observed in experiment. The enlargement on the right side of the panel includes further scattering vectors obtained from QPI at different sample areas or bias voltages (red circles), underlining the consistency of experimental data and theoretical modeling.
  }
  \label{fig:qpi}
\end{figure*}

The patches on the surface of Pb/Si(111) manifesting the $3 \times 3$ CO are too small to permit angle-resolved photoemission \cite{sup}. Instead, we use QPI \cite{QPI} to gain indirect access to the band structure. Quasiparticles elastically scattered at impurities will form a standing wave pattern in the differential tunneling conductance that is linked to the band structure of the examined material. The wavelength of these energy-dependent Friedel oscillations can be extracted via a Fourier transform. For energies within 50\,meV around the Fermi energy we recorded LDOS maps that reveal a distinct set of scattering vectors in their Fourier transformed images. A prototypical image for $V=+10$\,mV is shown in Fig.\ \ref{fig:qpi}(a). Although the scattering pattern seems fairly complex, it can be explained by a single scattering channel. As indicated by the red circles in the bottom right half of Fig.\ \ref{fig:qpi}(a), the whole pattern can be reconstructed by means of almost circular features centered around the Bragg peaks of the Fourier-transformed image. This means that in addition to the elastic scattering present in any QPI experiment there is a particularly strong contribution of quasiparticles scattered by an additional reciprocal lattice vector.
A closer analysis of measurements at different bias voltages reveals that the scattering vectors increase for more positive bias voltages, pointing towards an electronlike dispersion, {\it i.e.}, the ground state is a {\it metallic} CO. Interpolating the results from all energies and different sample spots yields a Fermi wave vector $k_f=(0.22 \pm 0.04)\,$\r{A}$^{-1}$, which agrees well with the value determined in Ref.\,\cite{Cren}.

\emph{Extended Hubbard model.---} For better theoretical insight into the STM results we model the Pb adatom system as an extended Hubbard model with isotropic hopping integrals $t_{ij}\equiv t_{|i-j|} \equiv t_n$ between the $n$th neighboring Pb atoms. We perform {\it ab initio} calculations to obtain accurate values for the hopping integrals \cite{sup}.
Local electron-electron interactions are included as a Hubbard on-site term with amplitude $U_0$ while nonlocal interactions are accounted for by the nearest-neighbor Coulomb term with amplitude $U_1$. The total Hamiltonian reads

\begin{equation}\begin{split}
  \mathcal{H} =& \sum_{ij,\sigma} \left( t_{ij} \,c_{i\sigma}^\dag c_{j\sigma}^\pd +
   {\rm H.c.}\right)   \\
  &+\, U_0 \sum_{i} n_{i\uparrow}n_{i\downarrow}+ U_1  \sum_{\langle ij\rangle}n_{i}n_{j}\ .  
\label{eq:ham}
\end{split}\end{equation}
Here $c_{i\sigma}^\dag$\,($c_{i\sigma}^\pd$) denotes a fermionic creation (annihilation) operator, $n_{i\sigma}\equiv c_{i\sigma}^\dag c_{i\sigma}^\pd$, and $n_i=n_{i\up}+n_{i\dw}$.
We consider a filling of one electron per lattice site, in agreement with our {\it ab initio} calculations.

\begin{figure}[t!]
  \centering
  \includegraphics[width=0.47\textwidth]{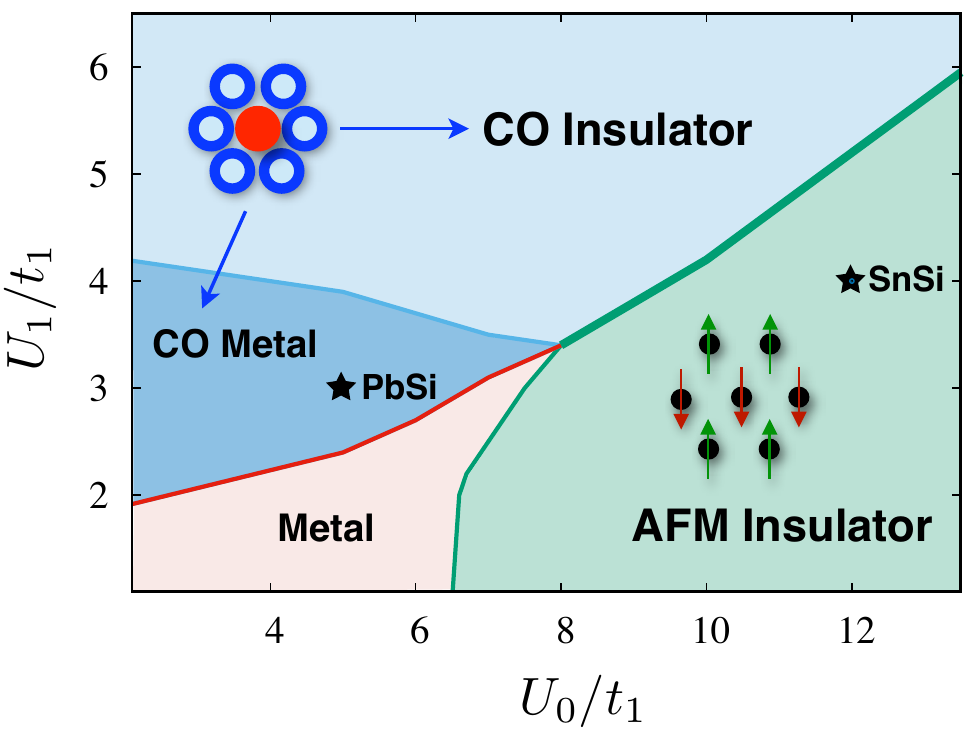}
  \caption{Quantitative phase diagram for the extended Hubbard Hamiltonian (Eq.\ \ref{eq:ham}) as obtained within XVCA. For dominant on-site-repulsion $U_0$ we find an antiferromagnetic insulator with row-wise order,  for sufficiently strong nearest-neighbor repulsion $U_1$ a charge ordered insulator. For weaker interactions also metallic regimes are present. Used parameters are $t_2/t_1=-0.383$, $t_3/t_1=0.125$ with $(U_0/t_1, U_1/t_1) = (5,3)$ for PbSi and (12,4) for SnSi. The energy scale for PbSi (SnSi) is set by $t_1=58.5$\,meV ($t_1=52.7$\,meV).}
  \label{fig:phasedia}
\end{figure}

However, values for $U_0$, $U_1$, etc.\ can usually not be obtained from pure {\it ab initio} methods; instead we will use them as system parameters to be fitted to the experimental data. For dominating on-site interaction $(U_{1}/U_{0} \ll 1)$ one would expect magnetic long-range order, provided that frustration caused by the triangular lattice, hoppings and interactions is not strong enough to trigger spin liquid physics. In contrast, for sufficiently large ratio $U_1/U_0$ one would rather expect some type of CO. We note that geometrical frustration not only suppresses magnetic ordering, but also affects CO through frustrated nearest-neighbor Coulomb interactions.

In the following, we apply the XVCA\,\cite{Potthoff2003b, sup} and map out the interacting $U_0$--$U_1$ phase diagram for single particle parameters $t_2/t_1=-0.383$ and $t_3/t_1=0.125$. Pb/Si(111) and its sister compound Sn/Si(111)\,\cite{GangLiNatComm} exhibit different hopping integrals; it turns out, however, that the ratios $t_2/t_1$ and $t_3/t_1$ are essentially identical (with negligible $t_4/t_1$) for both compounds allowing for a universal interacting phase diagram containing both the Pb/Si and the Sn/Si system (see Fig.\,\ref{fig:phasedia}).

The phase diagram of Hamiltonian (1) reveals metallic and insulating phases. As speculated earlier we indeed find different many-body ground states which either realize antiferromagnetic order (as in the Sn/Si system) or CO (as in the Pb/Si system, see Fig.\,\ref{fig:exp}) depending on the interaction parameters $U_0$ and $U_1$.
We did not find any signs of other magnetically or charge ordered phases in the parameter regime covered by Fig.\,\ref{fig:phasedia}. In the insulating regime, we find the first order phase transition from CO to antiferromagnetism (AFM) for $U_1/U_0 \approx 0.43$.

The XVCA method allows us to compute the $k$-resolved single-particle spectral function.
Simulated plots on the $(U_0, U_1)$ manifold compared with the experimental data
allow us to locate the position of Sn/Si(111) \cite{sup} and Pb/Si(111) in the phase diagram Fig.\ \ref{fig:phasedia}.
The best match between QPI and XVCA is obtained for interaction parameters $(U_0/t_1, U_1/t_1) = (5,3)$. The resulting spectral function along high symmetry directions is depicted in Fig.\,\ref{fig:qpi}(b). In the enlargement on the right side of this panel, we show a comparison between the relevant part of the theoretical spectral function and the scattering vectors found in QPI, demonstrating consistency over the whole energy range covered.

XVCA accurately accounts for nonlocal Coulomb interactions by exactly treating them within the cluster and in a variational scheme between the clusters; thus it represents a major advantage concerning the degree of ``material-realistic'' modeling.
By considering electronic correlations beyond the perturbative regime, the reconstructions observed in STM as well as the electronic properties of both Pb/Si(111) and Sn/Si(111) can be explained consistently. Our analysis demonstrates that the interplay of local and nonlocal electron-electron interactions is responsible for the observed CO in Pb/Si(111). We also investigated a Peierls distortion which might be caused by nesting as an alternative mechanism for CO, but the bare susceptibilities are not compatible with such reasoning---both in the absence and presence of spin-orbit coupling (SOC), see Supplemental Material \cite{sup}. In Ref.\ \cite{Cren} it was argued that the system can be properly explained through a combination of electronic correlations and SOC. While the relativistic DFT+U calculations can produce CO, they conflict with the experimentally observed ``1-down---2-up'' configuration. Moreover, DFT calculations cannot account for physics close to or beyond the Mott transition, and a unifying description of Pb/Si(111) and Sn/Si(111) is not feasible because the quantum many-body character of these systems is ignored. While SOC generates a fine splitting in the computed noninteracting bands of Pb/Si(111), it is not involved in the phase transition to the CO state \cite{Cren}.
In turn, our combination of DFT and XVCA simulations (neglecting SOC) correctly describes the experimental findings and provides insight how the interplay of $U_0$ and $U_1$ drives the CO, as anticipated in Ref.\ \cite{BiermannGWDMFT}.
Thus, SOC appears to play only a secondary role. We do not expect any qualitative changes in our XVCA results due to SOC for the CO regime. As a side note, in the insulating magnetic regime there might be additional phases related to SOC with spiral or skyrmionic order \cite{spirals}, yet being beyond the present scope.

In summary, we have unambiguously shown that the low temperature phase of Pb/Si(111) is a charge-ordered state. Many body simulations using the XVCA method have identified electronic longer-ranged correlations as the driving force of this state. A detailed analysis of STM and STS data furthermore revealed that the Pb atoms are arranged in a ``1-down---2-up'' fashion---in contrast to the previous understanding \cite{PbSiSTMintrinsic, Cren}. By comparison of XVCA calculated spectral functions with QPI measurements, we were able to pinpoint this material system on a phase diagram of correlation parameters $(U_0/t_1, U_1/t_1)$ which can serve as a map in the search for other, potentially exotic, ground states like unconventional superconductivity. The material system used here is captivating by its simplicity, i.e., the limited number of ingredients and the well-defined correlation parameters that are easy to control, in comparison to the multi-elemental alloys used for the study of high-temperature superconductivity thus far. Notably, correlated 2D triangular lattices are expected to yield either chiral singlet or chiral triplet superconductivity\,\cite{vojta1999,mackenzie-03rmp657,nandkishore-12np158,kiesel2013,Weber2006,sato-17rpp076501, Cao2017}; the latter is known to host exotic Majorana fermions bound to the vortex cores\,\cite{nayak-08rmp1083,leijnse2012}.

We acknowledge financial support from the DFG through the W\"urzburg-Dresden Cluster of Excellence on Complexity and Topology in Quantum Matter -- \emph{ct.qmat} (EXC 2147, project-id 39085490) as well as
through the Collaborative Research Centers SFB 1170 ``ToCoTronics'' and SFB 1143; S.R. acknowledges support from an Australian Research Council Future Fellowship (FT180100211).
S.R. and M.L. thank the Center for Information Services and High Performance Computing (ZIH) at TU Dresden for allocation of computer time. A.F. thanks the J\"ulich Supercomputing Centre for computer resources (project No hwb03).
\\

\end{document}